\documentclass[conference]{IEEEtran}
\usepackage[colorlinks,
linkcolor=red,
anchorcolor=green,
citecolor=blue
]{hyperref} 
\usepackage{amsfonts,amssymb,amsmath,array}
\usepackage{latexsym}
\usepackage{CJK}
\usepackage{cite}
\usepackage{bm}
\usepackage{color, soul}
\usepackage{colortbl}

\usepackage{graphicx}
\usepackage{epstopdf}
\usepackage{epsfig}
\usepackage{subfigure} 
\usepackage{framed}
\usepackage{verbatim}  
\usepackage{color,soul}
\definecolor{aliceblue}{rgb}{0.94, 0.97, 1.0}
\definecolor{blizzardblue}{rgb}{0.67, 0.9, 0.93}
\definecolor{antiquebrass}{rgb}{0.8, 0.58, 0.46}
\definecolor{beaublue}{rgb}{0.74, 0.83, 0.9}
\sethlcolor{blizzardblue}
\usepackage{mathrsfs}
\usepackage{algorithm} 
\usepackage{algorithmic} 
\usepackage{booktabs}
\usepackage{textcomp}
\usepackage{multirow}
\usepackage{lettrine}
\usepackage[scaled=0.92]{helvet}

\usepackage{algorithm}
\usepackage{algorithmic}

\usepackage{tcolorbox}

\usepackage{amsthm} 
\theoremstyle{remark} 

\ifCLASSINFOpdf
\else
\fi

\begin{document}
%
\title{A Switch to the Concern of User: Importance Coefficient in Utility Distribution and Message Importance Measure}
%
%
%

\author{\IEEEauthorblockN{Shanyun Liu\IEEEauthorrefmark{1},
Rui She\IEEEauthorrefmark{1},
Shuo Wan\IEEEauthorrefmark{1},
Pingyi Fan\IEEEauthorrefmark{1}, $Senior Member$, \textsl{IEEE},
and Yunquan Dong\IEEEauthorrefmark{2},
}
\IEEEauthorblockA{\IEEEauthorrefmark{1}
State Key Laboratory on Microwave and Digital Communications\\
Tsinghua National Laboratory for Information Science and Technology\\
Department of Electronic Engineering, Tsinghua University, Beijing,
P.R. China\\
Emails: \{liushany16, sher15, wan-s17\}@mails.tsinghua.edu.cn, fpy@tsinghua.edu.cn
}
\IEEEauthorblockA{\IEEEauthorrefmark{2}School of Electronic and Information Engineering,\\
Nanjing University of Information Science and Technology, Nanjing, P.R. China\\
Email: yunquandong@nuist.edu.cn
}
}

\maketitle

\begin{abstract}
This paper mainly focuses on the utilization frequency in receiving end of communication systems, which shows the inclination of the user about different symbols. When the average number of use is limited, a specific utility distribution is proposed on the best effort in term of fairness, which is also the closest one to occurring probability in the relative entropy. Similar to a switch, its parameter can be selected to make it satisfy different users' requirements: negative parameter means the user focus on high-probability events and positive parameter means the user is interested in small-probability events. In fact, the utility distribution is a measure of message importance in essence. It illustrates the meaning of message importance measure (MIM), and extend it to the general case by selecting the parameter. Numerical results show that this utility distribution characterizes the message importance like MIM and its parameter determines the concern of users. 
\end{abstract}

\begin{IEEEkeywords}
Utility distribution; Message importance measure; Importance coefficient; Large deviation theory; Information theory
\end{IEEEkeywords}

\IEEEpeerreviewmaketitle

\section{Introduction}
With the explosive growth of data, the problem of data analysis and processing becomes more and more important in big data and wireless communication \cite{chen2014big,bi2015wireless}. It is difficult for traditional data processing technology to deal with massive sets of data, and thus many literatures focused on the new methods to process the big data, such as \cite{pourkamali2017preconditioned}. Among them, \cite{fan2016message} discusses the problem with taking message importance into account. In fact, importance is a fundamental concept in communication, which is used in error correction coding \cite{sun2017unequal} and statistical testing \cite{abadie2018on}. On the base of this framework, we think every message has two fundamental attributes, i.e., the amount of information and the importance of the message \cite{liu2017non}. Similarly to the amount of information, which is measured by Shannon entropy, we also need a quantity to measure the message importance. 

 Message importance measure (MIM) was proposed to characterize the message importance in the case where small-probability events contain most of important information \cite{fan2016message}. From the viewpoint of information theory, \cite{fan2016message} defined parametric entropy, which can highlight the message importance of those events with relatively small occurring probabilities. Moreover, the discussion in \cite{she2017focusing} argues that the a specific event is focused on by choosing corresponding importance coefficient in parametric MIM. The information divergence in big data is introduced in \cite{she2017amplifying}, which can amplify information distance. Non-parametric MIM is defined in \cite{liu2017non}, and a new compressed code mode is proposed based on it. In this code, the unimportant information is abandoned voluntarily to compress data, while standard compressions form the compressed version by removing the redundancy of data, such as source coding. In fact, both parametric MIM and non-parametric MIM can be efficiently used in the minority subsets detection, communication theory, data compression, and hypothesis testing \cite{fan2016message,she2017focusing,liu2017non}.

Large deviation theory is one of fundamental theories in information theory, which is widely used in hypothesis testing \cite{Elements}. Based on it, the large deviation from the expected outcome is near $0$. However, it is different when the received data is used by human being. In large deviation theory, the empirical distribution in client side agrees with the original probability distribution of random variable. However, in actual use, human beings use the data according to practical requirements, and thus the actual utilization frequency may be different from the original probability distribution. For example, a meteorological station measures all kinds of meteorological parameters, such as temperature, humidity and PM2.5, and transfers them to the user. Obviously, the utilization frequency of the user is different from the measuring frequency. Moreover, the different user also likes different types of content, and thus this utilization frequency depends on the specific requirements of users. For example, the small-probability events is very useful and important in the minority subsets detection \cite{phua2010comprehensive,zieba2015counterterrorism,ando2006information} and big data compression \cite{liu2017non}. However, in support vector machine (SVM), one prefer the event with high probability \cite{vapnik1982estimation}. 
 
In this paper, we find a specific utility frequency on the best effort in term of fairness, when the average number of use is limited. Its parameter likes a switch that determines what types of content we focus on. Further, the form of this specific utility frequency is the same with MIM, and its properties agrees with what is discussed in the previous studies \cite{she2017focusing}.

The rest of this paper is organized as follows. Section \ref{sec:two} introduces the definition of utility distribution and solves the optimization problem to give its mathematical version. In Section \ref{sec:three}, the relationship between utility distribution and its parameter is discussed. Then, in Section \ref{sec:four}, we compare the MIM and utility distribution, and find they are equivalent. Some numerical results are presented to verify our results in Section \ref{sec:five}. Finally, we conclude the paper in Section \ref{sec:six}.

\section{Setup of Utility Distribution}\label{sec:two}
Let $X_1,X_2,X_3,...,X_n$ be i.i.d $\sim P(x)$. This sequence $X$ is from an alphabet $\mathcal{X}=\{a_1,a_2,...,a_{|\mathcal{X}|}\}$. We adopt the notation $\textbf{X}$ to denote this sequence. The probability $P(x)$ determines the transmission or storage strategy of sequence $X$. For example, the entropy rate of sequence $X$ depends on probability distribution, and the sequence can be transmitted reliably if entropy rate is less than the channel capacity \cite{Elements}. Although the information themselves may not have differences from the viewpoint of transmission or storage, they do have different utility for different receivers. In fact, after receiving the sequence, people prefer to use the information that they need rather than the one with the maximum probability. Thus, different events not only have different occurring frequency, but also have different utilization frequency. To describe it, we propose \textit{utility distribution} to characterize the utilization frequency of every event. 

Let $U'(x)$ denote the number of use of symbol $x$ ($x\in \mathcal{X}$). In fact, $U'(x)$ can be any value that satisfies the demands of human beings. In this paper, let $E$ be the set of the average number of use which is less than or equal to a constant $\beta'$, i.e.,
\begin{equation}\label{equ: PQ_condition1}
E=\{U: \sum_{a \in \mathcal{X}} {P(a)U'(a)}  \le \beta' \}.
\end{equation}
Let $\beta'=\beta \sum\nolimits_{a \in {\mathcal{X}}} {U'(a)} $, we obtain
\begin{equation}\label{equ: PQ_condition1}
E=\{U: \sum_{a \in \mathcal{X}} {P(a)} \frac{U'(a)}{ \sum\nolimits_{a \in {\mathcal{X}}} {U'(a)}}  \le \beta \}, 
\end{equation}
where $\sum\nolimits_{a \in {\mathcal{X}}} {U'(a)}$ gives the total number of use of all kinds, and $\beta$ is a scaling factor.

For convenience, we define $U(a)= \frac{U'(a)}{ \sum\nolimits_{a \in {\mathcal{X}}} {U'(a)}}$ as normalized dull utility frequency, and $\sum\nolimits_{a \in {\mathcal{X}}} {U(a)} =1$. Thus, $0\le U(a)\le 1$ for any $a \in \mathcal{X}$. 
In fact, these utility distributions that satisfy this condition have interesting properties, which will be illuminated in the following paper.

Furthermore, the difference between utilization frequency and occurring frequency shows the user's subjective impact on this sequence. When $U(a)/P(a)=1$, we think the symbol $a$ is used fairly. The symbol $a$ is overused when $U(a)/P(a)>1$, while the symbol $a$ is underused when $U(a)/P(a)<1$. In this paper, we expect to find the utility distribution on the best effort in term of fairness. That is, we find a specific utility distribution $U^*$ in $E$ that is closest to $P$ in the relative entropy. In fact, 
if the utilization frequency is equal to the occurring probability distribution (i.e., all the data has complete fair usage), this problem is equivalent to large deviation theory. Due to Sanov's theorem \cite{Elements}, $ \Pr \left( {{1 \over n}\sum\nolimits_{i = 1}^n {P({X_i})}  \le \beta}\right)=P^n(E)\le (n+1)^{|\mathcal {X}|}2^{-nD(U^*\parallel P)}$, where $U^*=\arg \mathop {\min }\limits_{U \in E} D(U\parallel P)$ and $D(U\parallel P)$ is relative entropy between distribution $U$ and $P$.

In addition, for all $a \in \mathcal{X}$,
\begin{flalign}\label{equ: PQ_condition2}
   \sum\limits_{a \in \mathcal{X}} {P(a)U(a)}  &\le \beta   \\
   1 - \sum\limits_{a \in \mathcal{X}} {P(a)U(a)}  &\ge 1 - \beta   \tag{\theequation a}\label{equ: PQ_condition2 a}\\
   \sum\limits_{a \in \mathcal{X}} {U(a)}  - \sum\limits_{a \in \mathcal{X}} {P(a)U(a)}  &\ge 1 - \beta   \tag{\theequation b}\label{equ: PQ_condition2 b}\\
   \sum\limits_{a \in \mathcal{X}} {(1 - P(a))U(a)}  &\ge \alpha  \tag{\theequation c}\label{equ: PQ_condition2 c}
\end{flalign}
where $\alpha=1-\beta$. Thus, the set $E$ is equal to $\{P: \sum_{a \in \mathcal{X}} {(1-P(a))U(a)}  \ge \alpha \}$. 

Thus, the utility distribution $U^*$ is the solution of the following optimization problem.
\begin{flalign}\label{equ:PQ1 optization}
\mathcal{P}: \,\, \arg \mathop {\min }\limits_U\,\,\, &D(U\parallel P)  \\
\textrm{s.t.}\,\,\,&\sum_{a \in \mathcal{X}} {(1-P(a))U(a)}  \ge \alpha  \tag{\theequation a}\label{equ:PQ1 optization a}\\
& \sum\limits_{a \in  \mathcal{X}} {U(a)}=1 \tag{\theequation b}\label{equ:PQ1 optization b}
\end{flalign}
Using Lagrange multipliers, we take
\begin{equation}\label{equ: Lag mul}
\begin{split}
&J(U)=\sum\limits_{a \in \mathcal{X}} {U(a)\ln{{U(a)} \over {P(a)}}} \\
 &\quad \quad \quad \quad \quad + \lambda \sum\limits_{a \in  \mathcal{X}} {(1-P(a))U(a)}  + \mu \sum\limits_{a \in  \mathcal{X}} {U(a)} .
\end{split}
\end{equation}
Differentiating with respect to $P^*(x)$, we get
\begin{equation}
 \ln U^*(x) + 1 - \ln P(x) + \lambda (1 - P(x)) + \mu  .
\end{equation}
Setting the derivative to 0, and we get $U^*(x)=P(x) e^{-\lambda(1-P(x))-\mu-1}$. Then substituting this in the constraint $\sum\nolimits_{a \in  \mathcal{X}} {U(a)}=1$, we get $e^{\mu+1}=\sum\nolimits_{a \in  \mathcal{X}} {P(a)e^{-\lambda (1-P(a))}}$. Hence,
\begin{equation}\label{equ:PQ_res}
U^*(x)=\frac{P(x) e^{-\lambda(1-P(x))}}{\sum\nolimits_{a \in  \mathcal{X}} {P(a)e^{-\lambda (1-P(a))}}}
\end{equation}
where the constant $\lambda$ is chosen to satisfy $\sum\nolimits_{a \in \mathcal{X}} {(1-P(a))U(a)}  \ge \alpha$.

\section{Discussion of Parameter}\label{sec:three}
Let $\varpi=-\lambda$ and $P=\{p_1,p_2,...,p_n\}$, and we obtain
\begin{equation}\label{equ:PQ_define}
U^*(x)=\frac{p_j e^{\varpi(1-p_j)}}{\sum_{i=1}^n {p_i e^{\varpi (1-p_i)}}}.
\end{equation}
Assume that there is unique minimum $p_{\min}$ and unique maximum $p_{\max}$ in distribution $P$. Therefore, we obtain $p_1<p_2 \le p_3 \le... \le p_{n-1}< p_n$, and $p_{\min}=p_1$ and $p_{\max}=p_n$. In addition, let $U^*=(u_1,u_2,...,u_n)$.

In fact, the parameter $\varpi$ has strong impact on the utility distribution $U^*$. When $\varpi \to + \infty$,
\begin{flalign}\label{equ:PQ_min1}
   u_1=&\mathop {\lim }\limits_{\varpi  \to  + \infty } {{{p_{\min }}{e^{\varpi (1 - {p_{\min }})}}} \over {\sum_{i = 1}^n {{p_i}{e^{\varpi (1 - {p_i})}}} }} \\
   =& \mathop {\lim }\limits_{\varpi  \to  + \infty } {{{p_{\min }}{e^{\varpi (1 - {p_{\min }})}}} \over {{p_{\min }}{e^{\varpi (1 - {p_{\min }})}} + \sum_{{p_i} \ne {p_{\min }}} {{p_i}{e^{\varpi (1 - {p_i})}}} }}  \tag{\theequation a}\\
  =&   \mathop {\lim }\limits_{\varpi  \to  + \infty } {1 \over {1 + \sum\nolimits_{{p_i} \ne {p_{\min }}} {{{{p_i}} \over {{p_{\min }}}}{e^{\varpi ({p_{\min }} - {p_i})}}} }}  \tag{\theequation b}\\ 
  =&   1 \tag{\theequation c} \label{equ:PQ_min1 c}
\end{flalign}
Obviously, $u_{k}=0$ when $k \ge 2$. Therefore, the utility distribution $U^*$ is $(1,0,0,...,0)$ in this case. 

In a similar way, we can get the utility distribution when $\varpi \to -\infty$. That is,
\begin{flalign}\label{equ:PQ_max1}
   u_n=&\mathop {\lim }\limits_{\varpi  \to  - \infty } {{{p_{\max }}{e^{\varpi (1 - {p_{\max }})}}} \over {\sum_{i = 1}^n {{p_i}{e^{\varpi (1 - {p_i})}}} }} \\
   =& \mathop {\lim }\limits_{\varpi  \to  - \infty } {{{p_{\max }}{e^{\varpi (1 - {p_{\max }})}}} \over {{p_{\max }}{e^{\varpi (1 - {p_{\max }})}} + \sum\limits_{{p_i} \ne {p_{\max }}} {{p_i}{e^{\varpi (1 - {p_i})}}} }} \tag{\theequation a} \\
  =&   \mathop {\lim }\limits_{\varpi  \to  - \infty } {1 \over {1 + \sum_{{p_i} \ne {q_{\max }}} {{{{p_i}} \over {{p_{\max }}}}{e^{\varpi ({p_{\max }} - {p_i})}}} }}  \tag{\theequation b}\\ 
  =&   1   \tag{\theequation c} \label{equ:PQ_max1 c}
\end{flalign}
Hence, $U^*=(0,0,..,0,1)$ in this case.

As mentioned above,  if the utilization frequency of a receiver is equal to the probability distribution (i.e., $U^*= P$), this problem is equivalent to large deviation theory. In this case, the parameter $\varpi$ have to satisfy
\begin{flalign}\label{equ: PQ_equ}
{p_j} &= {{{p_j}{e^{\varpi (1 - {p_j})}}} \over {\sum\limits_{i = 1}^n {{p_i}{e^{\varpi (1 - {p_i})}}} }}  
\end{flalign}
It is noted that $\varpi=0$ is the solution of (\ref{equ: PQ_equ}). Let $f(\varpi)={p_j} - {{{p_j}{e^{\varpi (1 - {p_j})}}} \over {\sum\nolimits_{i = 1}^n {{p_i}{e^{\varpi (1 - {p_i})}}} }}$ with respect to $\varpi$. Differentiate it with respect to $\varpi$, and we get
\begin{equation}\label{equ:diff_fw}
f'(\varpi)=-{{{p_j(1-p_j)}{e^{\varpi (1 - {p_j})}}} \over {\sum\nolimits_{i = 1}^n {{p_i(1-p_i)}{e^{\varpi (1 - {p_i})}}} }} \le 0.
\end{equation}
Thus, there is only one root for (\ref{equ: PQ_equ}), which is $\varpi=0$. In this case, $U^*=(p_1,p_2,...,p_n)$.

Due to $0 \le U(x) \le 1$, it is noted that
\begin{equation}
1-p_{\max} \le \sum_{a \in \mathcal{X}} {(1-P(a))U(a)} \le 1-p_{\min},
\end{equation}
 and thus we have
 \begin{flalign}
\Pr\left(\sum_{a \in \mathcal{X}} {(1-P(a))U(a)}  \ge 1-p_{\max} \right)&=1,  \\
\Pr\left(\sum_{a \in \mathcal{X}} {(1-P(a))U(a)}  > 1-p_{\min} \right)&=0. \tag{\theequation a}
\end{flalign}
According to $\alpha=1-\beta$, the average of utilization frequency is in $[p_{\min},p_{\max}]$. For comparison, the relationship between parameter and the utility distribution is summarized in Table I. 
\begin{table}[tbp]
\centering
    \caption{Table of utility distribution with parameters.}\label{tab:result1}
\begin{tabular}{c|c|c|c|c}
\toprule [1 pt]
$\varpi$ & $\lambda$ & $\alpha$ &$\beta$& $P^*$ \\
\hline
$-\infty$ & $+\infty$ & $1-p_{\max}$&$p_{\max}$& (0,0,...,0,1)\\
\hline
$0$ &  0 & $1-\sum\nolimits_{i = 1}^n {{p_i^2}} $ &$\sum\nolimits_{i = 1}^n {{p_i^2}} $& ($p_1,p_2,...,p_n$)\\
\hline
$+\infty$ &$-\infty$&  $1-p_{\min}$&$p_{\min}$ & (1,0,...,0,0)\\
\toprule [1 pt]
\end{tabular}
\end{table}

Actually, we obtain 
\begin{equation}
\sum_{i = 1}^n {{p_i^2}} = e^{-H_2(Q)}
\end{equation}
where $H_2(Q)$ is the R{\'{e}}nyi entropy $H_{\alpha}(\cdot)$ when $\alpha=2$ \cite{van2014renyi}. That is, when the user use the data according to the occurring probability, the average utilization probability shows the second order R{\'{e}}nyi entropy.

In fact, the utility distribution $U^*$ shows how people use the data. When $\varpi \to +\infty$, $U^*=(1,0,...,0)$, which means people prefer small-probability events. In this case, human beings take the high-probability events as granted or consider it as invalid information, and the data which they focus on and use is small-probability. For example, communication base station receives data, and this data usually involves many users' messages. As a result, for a particular user, the probability of message for him or her is small-probability and most data is useless. In this case,  the utilization frequency of this user is just like $(1,0,0...,0)$. Similarly, when $\varpi \to - \infty$, people would like to only focus on the high-probability events and consider the small-probability events as outliers which can be neglected. For example, we obey a special rule in SVM, which is that one only need to guarantee that the correct rate of algorithms is high-probability (not necessarily one) \cite{vapnik1982estimation}. The utilization frequency is equal to occurring frequency when $\varpi=0$. In this case, we do not take sides in any events. Let $X_1,X_2,...,X_N$ be a random sequence without human intervention, and the probability of condition $E$ is equal to $P^n(E)$. This problem is the large deviation theory in information theory, and \cite{Elements,sanov1958probability,csiszar1984sanov} discussed it. 

Parameter $\varpi$ is like a switch to determine users' interests. For $\varpi$ sufficiently large, people focus on the small-probability events, and the opposite is true when $\varpi$ approaches to negative infinity. Moreover, when absolute value of $\varpi$ is not large, the situation is complex, which depends on the form of occurring frequency and average utility frequency.

\section{Comparison between Utility Distribution and MIM} \label{sec:four}
Once parameter $\varpi$ is determined, $U^*$ depends only on the probability distribution of a random variable, which can be seen as an invariant of the system. In fact, the form of $U^*$ in (\ref{equ:PQ_define}) suggests that the user allocates utility proportion by weight factor $p_j e^{\varpi(1-p_j)}$. The process of using data can also be seen as the process of allocating processing resources. 
Assume that the symbol $X_i$ need processing resources with the size of $e^{\varpi (1-p_i)}$. Then, the average size of required processing resources is $\sum_{i=1}^n {p_i e^{\varpi (1-p_i)}}$, where the size of each symbol is $p_j e^{\varpi(1-p_j)}$. Therefore, the processing resources proportion of event $j$ is ${p_j e^{\varpi(1-p_j)}}/{\sum_{i=1}^n {p_i e^{\varpi (1-p_i)}}}$. 

In fact, the utility values are the subjective view of the user. They are not objective quantity in nature, and they only show the using tendency of the user. The event with larger utility values attract more people's interests. In other words, these utility values characterize the degree of importance of these events for the user. For example, one user focuses on the event with $a$ in alphabet $\mathcal{X}$. If the messages which satisfy $X=a$ are damaged, it will bring the huge loss or the interruption of the following proceedings for the fact that the user is only interested in these messages and expect to use them. Therefore, we think the utility characterizes message importance qualitatively, and $\sum_{i=1}^n {p_i e^{\varpi (1-p_i)}}$ describes the total message importance.

As a user's subjective concept, the values of utility or message importance make no sense. The most important thing for us is their relative size. For convenience, we take $p_j e^{\varpi(1-p_j)}$ as the measure of message importance for event $j$, and thus the total message importance is measured by $\sum_{i=1}^n {p_i e^{\varpi (1-p_i)}}$.

In fact, this measure of message importance is exactly the same with MIM in \cite{fan2016message}. MIM is proposed to measure the message importance in the case where small-probability events contains most valuable information and the parameter $\varpi$ in MIM is called \textit{importance coefficient}. The importance coefficient is always positive in MIM, which is consistent with the conclusion of this paper since MIM focuses on small-probability events. \cite{she2017focusing} discussed the selection of importance coefficient, and it pointed out that the event with probability $p_j$ becomes the principal component in MIM when $\varpi=1/p_j$. Since the same form, the utility distribution also agrees with this conclusion, which is shown in Fig. \ref{fig:fig3}.

Although MIM is proposed based on information entropy, it can also be given from the viewpoint of utility distribution. That is, MIM can also be seen as a utility distribution which is obtained on the best effort in term of fairness (i.e., it is most closest to probability distribution of the random variable in the relative entropy) when the average of utilization frequency is smaller than or equal to a constant. This conclusion confirms  the rationality of MIM in one aspect. 

\section{Numerical Results} \label{sec:five}
In this section, some numerical results are presented to validate the results in this paper.

\begin{figure}
  \centerline{\includegraphics[width=8.0cm]{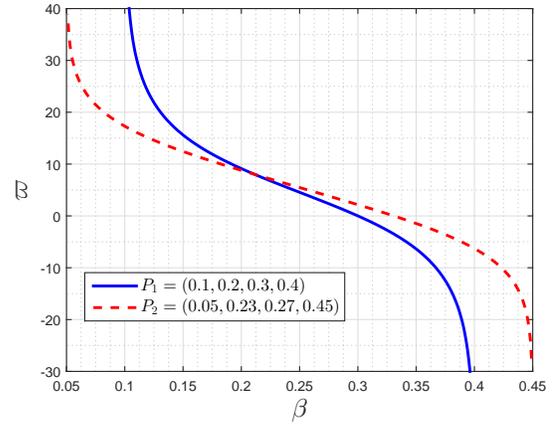}}
  \caption{$\beta$ vs $\varpi$.}\label{fig:fig1}
\end{figure}

Fig. \ref{fig:fig1} shows the relationship between the scaling factor of average utilization frequency $\beta$ and importance coefficient $\varpi$. The scaling factor of average utilization frequency $\beta$ is varying from $p_{\min}$ to $p_{\max}$, and the probability distribution $P_1=0.1,0.2,0.3,0.4)$ or $P_2=(0.05,0.23,0.27,0.45)$. 
It is noted that $\varpi$ decreases monotonically with the increasing of $\beta$. Moreover, there is a demarcation point $\beta_0$ ($\beta_0 \approx 0$) where both importance coefficient in $P_1$ and $P_2$ is equal. When $\beta$ is smaller than $\beta_0$, the importance coefficient in $P_1$ is larger than that in $P_2$, and the opposite is true when $\beta>\beta_0$. When $\beta \to p_{\min}$, we obtain $\varpi \to +\infty$, while $\varpi \to -\infty$ as $\beta \to p_{\max}$. When $\beta=\sum\nolimits_{i = 1}^n {{p_i^2}}$ ($0.1^2+0.2^2+0.3^2+0.4^2=0.3$ or $0.05^2+0.23^2+0.27^2+0.45^2=0.3308$), $\varpi=0$.

\begin{figure}
  \centerline{\includegraphics[width=8.0cm]{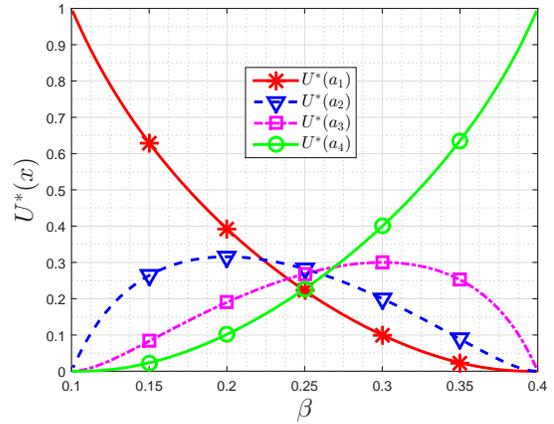}}
  \caption{$\beta$ vs $U^*$ when $P=(0.1,0.2,0.3,0.4)$.}\label{fig:fig2}
\end{figure}

\begin{figure}
  \centerline{\includegraphics[width=8.0cm]{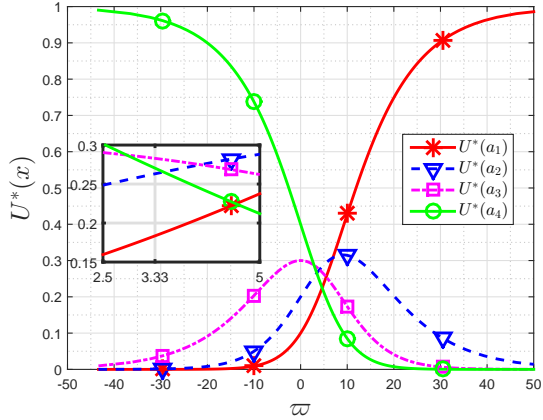}}
  \caption{$\varpi$ vs $U^*$ when $P=(0.1,0.2,0.3,0.4)$.}\label{fig:fig3}
\end{figure}
The probability distribution $P$ in Fig. \ref{fig:fig2} and Fig. \ref{fig:fig3} is $(0.1,0.2,0.3,0.4)$. $\beta$ is the scaling factor of average utilization frequency.

Fig. \ref{fig:fig2} shows the scaling factor of average utilization frequency $\beta$ versus the utility distribution $U^*$. $U^*(a_1)$ decreases monotonically with the increasing of $\beta$. In addition, $U^*(a_1)=1$ when $\beta=p_{\min}=p_1$ and $U^*(a_1)=0$ when $\beta=p_{\max}=p_4$. $U^*(a_j)$ ($j=2,3$) increases with the increasing of $\beta$ when $\beta<p_j$, and then it decreases with the increasing of $\beta$ when $\beta>p_j$. Therefore, $U^*(a_j)$ ($j=2,3$) achieves the maximum when $\beta=p_j$. Moreover, they are both $0$ when $\beta=p_{\max}$ or $\beta=p_{\min}$. $U^*(a_4)$ is opposite to $U^*(a_1)$. It increases in $(p_{\min},p_{\max})$, and $U^*(a_4)=1$ when $\beta=p_{\max}$ and $U^*(a_4)=0$ when $\beta=p_{\min}$.

Fig. \ref{fig:fig3} shows that importance coefficient $\varpi$ versus the utility distribution $U^*$. Fig. \ref{fig:fig3} is similar to the mirror image of Fig. \ref{fig:fig2} due to the fact that $\varpi$ decreases with the increasing of $\beta$. However, some other interesting observations are still obtained. $U^*\to(0,0,0,1)$ as $\varpi \to -\infty$ and $U^*\to(1,0,0,0)$ as $\varpi \to +\infty$. The utility distribution is equal to $P=(0.1,0.2,0.3,0.4)$ when $\varpi=0$. Moreover, we find that the utility of event with probability $p_j$ is larger than other three events' utility when $\varpi=1/p_j$. For example, when $\varpi=1/p_3=10/3$, $U^*(3)>U^*(j),\, j=1,2,4$. The utility of event with probability $p_j$ will be always less than one if $p_j\ne p_{\min}$ and $p_j\ne p_{\max}$.

\section{Conclusion}\label{sec:six}
In this paper, we discussed the problem of utility distribution, which reflects the concern of the user. Utility distribution is defined as the one that is the closest to original probability distribution in the relative entropy when the average number of use is limited. In fact, the utility distribution of each symbol is allocated by a special weight factor and this weight factor is a system invariant with a parameter, which can be seen as the measure of message importance. Moreover, the parameter, also called importance coefficient, determines the event type which people are interested in. When the importance coefficient is positive, the user focus on small-probability events, while the high-probability events attract users' interests when the importance coefficient is negative. In particular, as the importance coefficient approaches to positive infinity or negative infinity, users only concern the minimum probability event or the maximum probability event respectively. In particular, if utilization frequency exactly agrees with the occurring probability, the problem will be equivalent to the large deviation theory. In addition, the utility distribution is equal to parametric MIM due to the same form. The difference is that MIM focus on the situation where the small-probability events contain is more important and the utility distribution extends to general case. Discussing the applications of this utility distribution and analyzing different utility distribution in new restricted condition under wireless communication systems are of our future interests.

\section*{Acknowledgement}
This work was supported by National Natural Science Foundation of China (NSFC) NO. 61771283 and the China Major State Basic Research Development Program (973 Program) No.2012CB316100(2).


\end{document}